\documentclass[12pt]{article}

\usepackage{amssymb,amsmath,amsbsy,amsthm}
\usepackage{graphicx}
\usepackage{authblk}

\newcommand{\nr}[1]{{\bf {\it (#1)}}}
\newcommand{\BZ}{\mathrm{BZ}}

\newcommand{\pdag}{^{\phantom\dag}}

\newcommand{\Ref}[1]{(\ref{#1})}

\newcommand{\cE}{\mathcal{E}}

\newcommand{\cN}{\mathcal{N}}

\newcommand{\tV}{$t$-$V$}
\newcommand{\ta}{\tilde{a}}
\newcommand{\ee}{\,{\rm e}}
\newcommand{\ii}{{\rm i}}

\newcommand{\vn}{{\bf n}}

\newcommand{\vK}{{\bf K}}
\newcommand{\vk}{{\bf k}}
\newcommand{\ve}{{\bf e}}

\newcommand{\vp}{{\bf p}}

\newcommand{\Z}{{\mathbb Z}}

\newcommand{\HTC}{high-$T_c$ } 

\title{\Large{\bf{Fermions in two dimensions, bosonization, and exactly solvable models}}}
\date{\vspace{-1.0cm}\small June 9, 2012\vspace{0.2cm}}

\author[1,*]{Jonas de Woul}
\author[1,\dag]{Edwin Langmann}
\affil[1]{Department of Theoretical Physics, Royal Institute of Technology KTH\newline SE-106 91 Stockholm, Sweden \vspace{2mm}}

\begin{document}

\maketitle

\let\oldthefootnote\thefootnote
\renewcommand{\thefootnote}{\fnsymbol{footnote}}
\footnotetext[1]{Electronic address: {\tt jodw02@kth.se}}
\footnotetext[2]{Electronic address: {\tt langmann@kth.se}}
\let\thefootnote\oldthefootnote

\vspace{-1.5cm}

\begin{abstract}
We discuss interacting fermion models in two dimensions, and, in particular, such that can be solved exactly by bosonization. One solvable model of this kind was proposed by Mattis as an effective description of fermions on a square lattice. We review recent work on a specific relation between a variant of Mattis' model and such a lattice fermion system, as well as the exact solution of this model. The background for this work includes well-established results for one-dimensional systems and the high-$T_c$ problem. We also mention exactly solvable extensions of Mattis' model. 
\end{abstract}

\newcommand{\vv}{V}

\section{Introduction}
The theory of metals developed in the 1950s--60s is the basis of an excellent qualitative description of many materials that exist in nature. This theory is based on the Fermi liquid picture of nearly-free quasiparticles and standard methods like diagrammatic perturbation theory; it is thoroughly discussed in most introductory textbooks on condensed matter physics (see e.g.\ Ref.~\cite{Mahan}). 

Still, this otherwise so successful theory is known to fail for systems with strong correlation effects, as is typically observed in low-dimensional systems. For example, in the case of fermions confined to one dimension a different picture has been developed that radically departs from that of a Fermi liquid. This picture is largely based on prototype models that can be solved exactly by analytical techniques, and which nevertheless allow to take into account certain types of correlation effects. One important such technique is known as {\em bosonization}, which amounts to rewriting interacting fermions in terms of free bosons; we will return to this in later sections. For fermions confined to two dimensions, an equally satisfactory description has not yet been achieved. In this review, we describe a particular approach that uses a combination of mean field theory and bosonization to study such fermion systems~\cite{EL0,EL1,dWL1,dWL2,dWL3,dWL4}. In particular, we discuss a class of 
two-dimensional fermion models that can be solved exactly by bosonization, as well as their relation to models describing fermions on a square lattice. We also discuss the so-called \HTC problem providing one physical motivation for this work. 

Our plan is as follows. Section~\ref{sec1.1} introduces the models playing a central role in this paper, and Section~\ref{sec1.2} contains a biased overview of the literature on bosonization in higher dimensions.\footnote{The literature on higher dimensional bosonization is huge, and our discussion is far from exhaustive.} Section~\ref{sec2} contains a discussion of lattice fermion models and their relation to the \HTC problem (Section~\ref{sec2.1}). As a pedagogical way of introducing the results and notation used in later parts, we also discuss the bosonization approach to one dimensional lattice fermion systems (Sections~\ref{sec2.2}--\ref{sec2.3}).  Section~\ref{sec3} reviews results for a simple prototype two dimensional lattice fermion  system obtained in Refs.~\cite{EL0,EL1,dWL1,dWL2}. Section~\ref{sec4} describes extensions of the latter results to more complicated models~\cite{dWL3,dWL4}. The final Section~\ref{sec5} contains a somewhat subjective discussion of our results from a broader perspective. 
 
\subsection{Exactly solvable QFT models}
\label{sec1.1} 
The Luttinger model~\cite{Tomonaga,Thirring,Luttinger,MattisLieb} is a prominent example of an exactly solvable quantum field theory (QFT\footnote{Note that by "QFT" we mean a quantum model with infinitely many degrees of freedom.}) model that has become a prototype for correlated fermion systems in one spatial dimension (1D)~\cite{Haldane}. This model describes two flavors of fermions, labeled by an index $r=\pm$ and by 1D momenta $k$, and is defined by the Hamiltonian
\begin{equation}
\label{Luttinger} 
H_L = \sum_{r=\pm} \int dk\, v_F rk :\!\hat\psi_r^\dag(k)\hat\psi^{\pdag}_r(k)\!: + \sum_{r,r'=\pm}\int\frac{dp}{2\pi}\hat\vv_{r,r'}(p)\hat J_r(-p)\hat J_{r'}(p)
\end{equation} 
with $\hat\psi^{(\dag)}_r(k)$ fermion field operators defined by the usual canonical anti-commutator relations (CAR), 
\begin{equation}
\hat J_r(p) =  \int dk\,:\!\hat\psi_r^\dag(k)\hat\psi_r^{\pdag}(k+p)\!: 
\end{equation} 
fermion density operators, $\hat\vv_{r,r'}(p)$ (suitable\footnote{See e.g.\ \cite{HSU}.}) two-body potentials, $v_F>0$ the Fermi velocity, and the colons indicating normal ordering. The technique to solve this model is based on mathematical results collectively known as bosonization; see \Ref{fact1}--\Ref{fact3} below. It is worth emphasizing that ``exactly solvable'' has a very strong meaning for this model: not only the eigenstates, but also all correlations functions can be computed analytically; see Ref.\ \cite{HSU} and references therein.

As first pointed out by Mattis~\cite{Mattis}, there exists a similar model describing fermions in 2D and which is also exactly solvable by bosonization. This model describes four flavors of fermions that are labeled by two indices $r,s=\pm$ and by 2D momenta $\vk=(k_1,k_2)$; it is defined by the Hamiltonian
\begin{equation} 
\label{Mattis' Hamiltonian}
\begin{split} 
H_M =& \sum_{r,s=\pm} \int d^2k\, v_F r k_s  :\!\hat\psi_{r,s}^\dag(\vk)\hat\psi_{r,s} (\vk)\!: \\ &+ \sum_{r,s,r',s'=\pm}\int\frac{d^2p}{(2\pi)^2}\hat\vv_{r,s,r',s'}(\vp)\hat J_{r,s}(-\vp)\hat J_{r',s'}(\vp), 
\end{split}
\end{equation} 
with $k_s=(k_1+sk_2)/\sqrt{2}$, and $\hat J_{r,s}(\vp) =  \int d^2k\,:\!\hat\psi_{r,s}^\dag(\vk)\hat\psi_{r,s}^{\pdag}(\vk+\vp)\!: $ etc., similarly as above. Mattis, having the famous \HTC problem of the cuprate superconductors~\cite{BednorzMueller} in mind, proposed this model as an effective description of fermions on a square lattice. We believe that the Mattis model deserves to be better known: First, exactly solvable models in 2D are rare, and any physically motivated such example deserves to be studied from different points of view. Second, as proposed and elaborated by us in recent work~\cite{EL0,EL1,dWL1,dWL2}, the exact solution of this model is a key part of a method to compute physical properties of 2D lattice fermion systems. Third, there exist exactly solvable extensions of the Mattis model that allow, for example, to study the effect of dynamical electromagnetic fields and spin on 2D interacting fermion systems by exact solutions~\cite{dWL3,dWL4}.

A main part of this review is on the relation of the Mattis model and a model of fermions on a square lattice. This relation is based on a method to derive effective QFT models for such lattice fermion systems. This method, which we call {\em (partial) continuum limit}, is well-established in 1D, and in the simplest case of spinless lattice fermions leads to the Luttinger model. The extension of this method to spinless fermions on a square lattice leads to a model that, in addition to the fermion fields $\hat\psi^{(\dag)}_{r,s}(\vk)$, $r,s=\pm$, that describe the so-called {\em nodal} fermion degrees of freedom, also takes into account the so-called {\em antinodal} fermion degrees of freedom represented by operators $\hat\psi^{(\dag)}_{r,0}(\vk)$, $r=\pm$. While the energy dispersion relations of the nodal fermions are linear in the momenta, $\epsilon_{r,s}(\vk)=v_F rk_s$ for $r,s=\pm$, the antinodal fermions have hyperbolic dispersion relations, $\epsilon_{r,0}(\vk)=rc_Fk_+k_- -\mu_0$ for $r=\pm$, with computable constants $c_F$ and $\mu_0$. The full Hamiltonian providing an effective description of 2D lattice fermions is the following extension of the Mattis model 
\begin{equation} 
\label{2DLutt} 
\begin{split} 
H =& \sum_{r,s=\pm} \int d^2k\, v_F r k_s  :\!\hat\psi_{r,s}^\dag(\vk)\hat\psi_{r,s} (\vk)\!: \\ & + \sum_{r=\pm} \int d^2k\, (c_F r k_+k_--\mu_0)  :\!\hat\psi_{r,0}^\dag(\vk)\hat\psi_{r,0} (\vk)\!:  \\ &+ \sum_{r,r'=\pm}\sum_{s,s'=0,\pm} \int\frac{d^2p}{(2\pi)^2}\hat\vv_{r,s,r',s'}(\vp)\hat J_{r,s}(-\vp)\hat J_{r',s'}(\vp)
\end{split}
\end{equation} 
with two-body interaction potentials $\hat\vv_{r,s,r',s'}$ determined by the underlying lattice fermion interactions~\cite{EL1}. The model defined by the Hamiltonian in \Ref{2DLutt} is a 2D analogue of the Luttinger model in the sense that it arises as a partial continuum limit of 2D lattice fermions, and that it is amenable to bosonization. We emphasize that this 2D Luttinger model is not exactly solvable; only the nodal fermion degrees of freedom can be mapped exactly to non-interacting bosons and thus be treated without approximations. Treating the antinodal fermions by mean field theory, one finds a significant parameter regime away from half filling for which the antinodal fermions are gapped~\cite{dWL1}. We proposed that, in this partially gapped phase, the low-energy physics of the system can be described by the Mattis model~\cite{dWL1}. This motivated us to study the Mattis model from a mathematical point of view, and we found that the Mattis model is indeed exactly solvable in the same strong sense as the Luttinger model~\cite{dWL2}. 

\subsection{Bosonization in higher dimensions (review)} 
\label{sec1.2}
The literature on bosonization in dimensions higher than one is quite extensive. Below we mainly discuss work whose bosonization methods lie closest to ours, while references to other approaches are briefly mentioned at the end. 
 
The first to apply bosonization to interacting, non-relativistic fermions in higher dimension was Luther~\cite{221} (see also Ref.~\cite{HSU} for contemporary work). For a discretized spherical Fermi surface in three dimensions, one can bosonize in the radial direction at each point on the surface, while treating the discretized angles as flavor indices~\cite{221}; this is sometimes called radial- or tomographic bosonization. Using a generalized Kronig identity, one can then express the kinetic part of the Hamiltonian in terms of densities~\cite{221,HSU}. Similarly, the fermion fields can be written in terms of charge shift operators (also called Klein factors), which depend on the radial direction, and an exponential of boson operators~\cite{HSU}. However, in this early approach, only density operators with momenta in the radial direction behave as bosons, and thus interactions with transverse momentum exchange cannot be treated exactly. 

Real interest in higher-dimensional bosonization came with the \HTC problem and the prospect of finding non-Fermi-liquid behaviour in strongly correlated 2D models. In particular, Anderson suggested that the 2D Hubbard model on a square lattice has a Luttinger liquid phase away from half-filling, and that this could be explored using bosonization methods~\cite{222}. Whether or not Luttinger-liquid behaviour is possible in this model is still an open problem. Rigorous work on the renormalization group have shown that weakly-coupled 2D fermions with Fermi surfaces of non-zero curvature are in general Fermi liquids~\cite{223} (see also Ref.~\cite{7} and references therein). Most of the attention has therefore been on Fermi surfaces with ``flat parts'', i.e. they contain portions that are straight (see for example Refs.~\cite{225,226}). 

Prior to Anderson's suggestion, Mattis proposed a 2D model of fermions with density-density interactions that could be solved exactly using bosonization~\cite{Mattis}. The kinetic part contained four types of fermions with linearised tight-binding band relations on each side of a square Fermi surface (see \Ref{Mattis' Hamiltonian}), which is reasonable for a Hubbard-type model with short-range hopping and near half-filling. Unlike Ref.~\cite{221}, the density operators were taken to be bosonic for all momentum exchange, although no details were provided. Rewriting the kinetic part in terms of densities, the Hamiltonian of the model could be diagonalized using a Bogoliubov transformation.

In more recent work, Luther~\cite{228} also studied fermions on the square Fermi surface with linear tight-binding band relations. Let $k_\parallel$ and $k_\perp$ denote the momenta parallel and perpendicular to a face of the square Fermi surface. Following Ref.~\cite{221}, one would treat $k_\parallel$ as a flavor index, extend $k_\perp$ to plus and minus infinity, and introduce a Dirac sea in which all states $k_\perp<0$ are filled. However, unlike the 1D case, one should note that there is a huge degeneracy in choosing the accompanying flavor index to the unbounded momenta. One can for example do a Fourier transformation (change of basis) in the $k_\parallel$-direction and then bosonize the fermions with a new flavor index $x_\parallel$. These can be interpreted as coordinates for a collection of parallel chains aligned in the $k_\perp$-direction. In this way, Luther was able to include boson-like interactions with momentum exchange also in the $k_\parallel$-direction~\cite{228}. The properties of this model were further investigated in Refs.~\cite{229,230}.

Finally, we mention other work on rewriting interacting fermions as bosons, although these methods are somewhat different from that discussed here. Haldane~\cite{231} formulated a phenomenological approach in which density fluctuations corresponding to momentum exchange of states near an arbitrary $D$-dimensional Fermi surface are assumed bosonic. These ideas were further pursued in the works of Refs.~\cite{232,235,242}. Bosonization methods using functional integration and Hubbard-Stratonovich transformations have been developed in Refs.~\cite{245,248}. There is also an approach based on non-commutative geometry~\cite{252}.

\section{Motivation and background} 
\label{sec2} 
\subsection{High-$T_c$ problem and 2D lattice fermions}
\label{sec2.1} 
The possible violation of Landau's Fermi liquid theory in models of strongly interacting fermions has been an actively researched problem for many years. Interest in this topic quickly grew with the discovery of high temperature superconductivity in the cuprates~\cite{BednorzMueller} and the realization that these materials display many properties not described by Fermi liquid theory; see e.g.\ Ref.\ \cite{Bonn} for review. Early on, it was suggested that 2D lattice fermion models of Hubbard-type capture the strongly correlated physics of cuprates~\cite{AndersonRVB,Emery,VSA,ZhangRice}. These models are appealing due to their apparent simplicity. For example, the much-studied 2D Hubbard model can be defined by the  Hamiltonian
 \begin{equation}
 \label{HtU}
 H_{tU} = -t\sum_{\langle i,j\rangle}\sum_{\alpha=\uparrow,\downarrow} \left(c^\dag_{i,\alpha} c^{\pdag}_{j,\alpha} +c^\dag_{j,\alpha} c^{\pdag}_{i,\alpha}\right) + U\sum_{i}n_{i,\uparrow}n_{i,\downarrow},\quad n_{i,\alpha}\equiv c^\dag_{i,\alpha} c^{\pdag}_{i,\alpha} 
 \end{equation} 
with fermion operators $c^{(\dag)}_{i,\alpha}$ labeled by sites $i$ of a 2D lattice and a spin index $\alpha=\uparrow,\downarrow$, and $\sum_{\langle i,j\rangle}$ a sum over all nearest-neighbor (nn) pairs on this lattice ($t>0$ and $U>0$ are the usual Hubbard parameters). The Hubbard model is conceptually simple since, if one restricts to a lattice with a finite number of sites, the Hamiltonian in \Ref{HtU} can be represented by a finite, albeit usually large, matrix\footnote{For a lattice with $\cN$ sites, the size of this matrix is $4^\cN\times 4^\cN$.}.
   
However, it has proven to be very difficult to do reliable computations for 2D Hubbard-like models for lattice sizes and intermediate coupling values of interest for the cuprates (i.e.\ for $U/t$ in the range between 2 and 10, say). Thus, despite of much work over many years, no consensus has been reached on the physical properties of such models in large parts of the interesting parameter regime. To be more specific: one important parameter for Hubbard-like systems is {\em filling} $\nu$, which is defined as the average fermion number per lattice site. For the Hubbard model this parameter is in the range $0\leq\nu\leq 2$. There is consensus that, at half-filling (i.e.\ for $\nu=1$), the 2D Hubbard model describes a Mott insulator. 

The challenge is to study the 2D Hubbard model close to, but away from, half filling, which is a regime that is much less understood. Our approach has been to look for other models that describe the same low-energy physics but are more amenable to reliable computations. As we explain in Section~\ref{sec3.1}, our search for such models was motivated by mean field results for the 2D Hubbard model~\cite{LW}. Another important motivation and guide were 1D lattice fermion systems that, different from 2D, are well understood also by numerical and analytical methods: In Sections~\ref{sec2.2}--\ref{sec2.3} we review a particularly useful method in 1D, namely to perform a suitable continuum limit to obtain a QFT model that can be studied by bosonization. As already mentioned, this method is well-established in 1D, but we present it in a way that makes our generalization to 2D~\cite{EL1} natural. For simplicity, much of our discussion in this paper is on a spinless variant of the Hubbard model defined by the Hamiltonian
\begin{equation}
\label{HtV} 
H_{tV} = -t\sum_{\langle i,j\rangle}\left(c^\dag_i c^{\pdag}_j+c^\dag_j c^{\pdag}_i\right)  + \frac{V}{2} \sum_{\langle i,j\rangle} n_in_j, \quad n_i\equiv c^\dag_i c^{\pdag}_i
\end{equation} 
with fermion operators $c^{(\dag)}_i$ labeled only by lattice sites $i$; $t>0$ and $V>0$ are model parameters. For this so-called \tV model, the filling parameter is in the range $0\leq \nu\leq 1$. Similarly as the Hubbard model, the 2D \tV model is believed to describe an insulator at half-filling $\nu=1/2$ and intermediate coupling values $V/t>0$, and it has been a challenge to understand the model away from half-filling. We will also shortly describe generalizations of our results for the 2D \tV model to the Hubbard model in Section~\ref{sec4.2}. 

\subsection{1D lattice fermions and the Luttinger model}
\label{sec2.2} 
We now describe a method to derive the Luttinger model from the 1D variant of the so-called \tV model defined in \Ref{HtV}. 

In 1D, the lattice sites are $i=1,2,\ldots,\mathcal{N}$. Introducing a lattice constant $a>0$ and the "length of space" $L\equiv \mathcal{N}a$,  we define $\psi(x)\equiv a^{-1/2}c_j$ with the 1D spatial positions $x=ja$. Due to the finite $L$, possible fermion momenta $k$ are in the set $\Lambda^*\equiv (2\pi/L)(\Z+1/2)$ (we use anti-periodic boundary conditions). Using lattice Fourier transform, we express the Hamiltonian in \Ref{HtV} in terms of fermion operators $\hat\psi^{(\dag)}(k)$ on the Brillouin zone of the lattice\footnote{We use anti-periodic boundary conditions.} 
\begin{equation}
\label{BZ}
\BZ = \left\{ k\in\Lambda^*; \quad -\frac{\pi}{a}<k<\frac{\pi}{a}\right\} 
\end{equation} 
as follows
\begin{equation} 
\label{HtV1D} 
\begin{split} 
H_{tV} &= \int_{\BZ}  \hat{d}k\, \epsilon(k) \hat\psi^\dag(k)\hat\psi(k) \\ &+ \int_{\BZ} \hat{d}k_1\cdots  \int_{\BZ} \hat{d}k_4 v(k_1,\cdots,k_4) 
 \hat\psi^\dag(k_1)\hat\psi(k_2)  \hat\psi^\dag(k_3)\hat\psi(k_4) 
 \end{split} 
 \end{equation} 
with the {\em band relation} $\epsilon(k)=-2t\cos(ka)$ and the {\em interaction vertex}  
\begin{equation} 
v(k_1,\ldots,k_4) = \hat u(k_1-k_2)\sum_{n\in\Z}\hat\delta(k_1-k_2+k_3-k_4+(2\pi/a) n )
\end{equation} 
where $\hat u(p)=aV\cos(ap)/(2\pi)$;  we write  $\int_S \hat dk\equiv \sum_{k\in S}(2\pi/L)$ for scaled Riemann sums ($S$ some subset of $\Lambda^*$)  and  $\hat\delta(k_1-k_2)\equiv [L/(2\pi)]\delta_{k_1,k_2}$ for scaled Kronecker deltas. Our normalization of the CAR is such that $\{\hat\psi(k_1),\hat\psi^\dag(k_2)\} = \hat\delta(k_1-k_2)$. We note that these scalings are such that all formulas remain meaningful in the formal infinite volume limit $L\to\infty$, but it is convenient to keep $L$ finite during computations.  We use a particle physics jargon and refer to $a$ and $L$ as {\em UV- and IR-cutoff}, respectively. 

Our aim is to modify the model so as to make it better amenable to analytical computations but not (much) change the low-energy properties of the model. The strategy is to perform a continuum limit $a\to 0^+$, i.e., the lattice constant $a$ is a small parameter, and sub-leading terms in $a$ are assumed to be of lesser importance. However, this limit has to be taken with care: Physics folklore suggests that the low-energy physics of a fermion system is dominated by momentum states close to the Fermi surface, i.e., one should only change the Hamiltonian so that only momentum states are affected that are far away from the Fermi surface. To be more specific: In the non-interacting case $V=0$, the groundstate of the Hamiltonian in \Ref{HtV1D} is 
\begin{equation}
|\Omega\rangle = \prod_{k:\, \epsilon(k)-\mu\leq 0}\hat\psi^\dag(k)|0\rangle
\end{equation} 
with the chemical potential $\mu$ determined by the fermion density and $|0\rangle$ the vacuum state such that $\hat\psi(k)|0\rangle =0$ for all $k$. We now choose $Q$ such that $\epsilon(rQ/a)=\mu$ for $r=\pm$, i.e., the filled momentum states are in the range $-Q/a<k<Q/a$, and the two points $\pm Q/a$ correspond to the Fermi surface. Note that filling in this situation is $\nu=Q/\pi$, i.e., half-filling corresponds to $Q=\pi/2$. The groundstate $|\Omega\rangle$ is then fully characterized by the following conditions,
\begin{equation}
\label{Dirac} 
\hat\psi_\pm(\pm k) |\Omega\rangle = \hat\psi^\dag_\pm(\mp k) |\Omega\rangle = 0\quad \forall k>0
\end{equation}  
where $\hat\psi_r(k)\equiv \hat\psi(K_r+k)$, $r=\pm$, are fermion operators labeled my momenta close to the Fermi surface points $K_r\equiv rQ/a$. 

To prepare for this limit it is convenient to rewrite the \tV model as a model of two flavors of fermions. For that we divide the Brillouin zone in two subsets $\Lambda^*_r=\{ k\in\BZ-K_r; r(k+K_r)>0 \} $ such that
\begin{equation}
\int_{\BZ}\hat d k\, f(k) = \sum_{r=\pm} \int_{\Lambda^*_r}\hat d k\, f(K_r+k)
\end{equation} 
for arbitrary functions $f$ defined on the Brillouin zone. This shows that the fermion operators $\hat\psi_r(k)$ are to be labeled by momenta $k\in\Lambda^*_r$, $r=\pm$. Inserting this in \Ref{HtV1D} we obtain a Hamiltonian describing two flavors of fermion operators $\hat\psi_r(k)$, $k\in\Lambda^*_r$ and $r=\pm$, with dispersion relations $\epsilon_r(k)\equiv \epsilon(K_r+k)$ and interaction vertices
\begin{equation}
\label{1Dvertex} 
\begin{split} 
&v_{r_1,\cdots,r_4}(k_1,\ldots,k_4) =  \hat u(K_{r_1}-K_{r_2}+k_1-k_2)\\ & \times \sum_{n\in\Z}\hat\delta(K_{r_1}-K_{r_2} + K_{r_3}-K_{r_4} + k_1-k_2+k_3-k_4+(2\pi/a) n ),
\end{split} 
\end{equation} 
i.e., $H_{tV} = \sum_{r=\pm} \int_{\Lambda^*_r}  \hat{d}k\, \epsilon_r(k) \hat\psi_r^\dag(k)\hat\psi_r(k) +\cdots$ (the dots indicate obvious interaction terms). 

We now modify the model as follows: First, we Taylor expand the dispersion relations $\epsilon_r(k)=\mu + rv_F k +O(a^2k^2)$, $v_F=2ta\sin(Q)$, and ignore the higher order terms, i.e., we replace
\begin{equation}
\label{app1} 
\epsilon_r(k)\to \mu + rv_F k. 
\end{equation} 
Second, we replace the interaction vertex by 
\begin{equation}
\label{app2} 
\begin{split} 
v_{r_1,\cdots,r_4}(k_1,\ldots,k_4) \to  \hat u(K_{r_1}-K_{r_2})\hat\delta(k_1-k_2+k_3-k_4) 
\\ \sum_{n\in\Z}\hat\delta( K_{r_1}-K_{r_2} + K_{r_3}-K_{r_4} + (2\pi/a) n ), 
\end{split} 
\end{equation} 
i.e., we expand $\hat u(K_{r_1}-K_{r_2} +k_1+k_2)=\hat u((r_1-r_2)Q/a)[1+O(a(k_1-k_2))]$ and ignore the higher order terms and, at the same time, modify allowed scattering terms in the interaction but only those where $a(k_1-k_2+k_3-k_4)$ is "large".  Since momenta $k$ close to the Fermi surface are such that $ka$ is "small", we expect that these changes are appropriate. 

Inserting these changes in the Hamiltonian one obtains, after some computations,
\begin{equation} 
\label{Lutt1}
H_{tV} \to \sum_{r=\pm} \int_{\Lambda^*_r}  \hat{d}k\, rv_F k \hat\psi_r^\dag(k)\hat\psi_r(k) + \int_{\tilde\Lambda^*} \frac{\hat{d}p}{2\pi} \, \hat v(p) \hat\rho_+(p)\hat\rho_-(-p) 
\end{equation}  
with
\begin{equation} 
\label{rho} 
\hat\rho_r(p) =  \int_{\Lambda^*_r}  \hat{d}k\, \hat\psi_r^\dag(k)\hat\psi_r(k+p)  
\end{equation} 
and\footnote{$\theta$ below is the Heaviside function.}
\begin{equation} 
\label{hatv} 
\hat v(p) = g\theta(\pi-|p|a) 
\end{equation} 
with $g=2aV\sin^2(Q)/\pi$; here and in the following we use the notation $\tilde\Lambda^*\equiv(2\pi/a)\Z$. We note that, in these computations, one can ignore additive constants and chemical potential terms, i.e.\  terms proportional to $\sum_{r=\pm} \int_{\Lambda^*_r}  \hat{d}k\, \hat\psi_r^\dag(k)\hat\psi_r(k) $ (since the former terms amount to constant shifts of the energy, and the latter to a renormalization of the chemical potential, which both are  irrelevant). The final modification is to  replace on the r.h.s.\ in \Ref{Lutt1} and in \Ref{rho} 
\begin{equation} 
\label{app3} 
\Lambda_r^*\to \Lambda^*
\end{equation} 
i.e., to remove the restriction on fermion momentum states by taking the limit $a\to 0^+$ in the sets $\Lambda^*_r$, $r=\pm$. We note that it is this approximation which turns the model into a QFT: before we had a model with a finite number of fermion momentum states, but after \Ref{app3} we have a quantum model with an infinite number of degrees of freedom.  One finds that, in order to obtain a model that is mathematically well-defined, one needs to normal order {\em before} making the change in \Ref{app3}: defining 
\begin{equation} 
:\!A\!:\, \equiv A-\langle\Omega,A\Omega\rangle 
\end{equation} 
for operators $A$ and with the state $|\Omega\rangle$ fully characterized by \Ref{Dirac}, one can show that 
\begin{equation} 
\hat J_r(p) \equiv \int_{\Lambda^*}  \hat{d}k\, :\! \hat\psi_r^\dag(k)\hat\psi_r(k+p)\!:  
\end{equation} 
and 
\begin{equation} 
\label{Lutt2} 
H_L \equiv \sum_{r=\pm} \int_{\Lambda^*}  \hat{d}k\, rv_F k :\!\hat\psi_r^\dag(k)\hat\psi_r(k)\!: + \int_{\tilde\Lambda^*} \frac{\hat{d}p}{2\pi} \, \hat \vv(p) \hat J_+(p) \hat J_-(-p)   
\end{equation} 
are mathematically well-defined. Note that normal ordering, before the replacement in \Ref{app3}, only amounts to dropping irrelevant additive constants and chemical potential terms.  

The Hamiltonian in \Ref{Lutt2} makes mathematically precise a special case of the one in \Ref{Luttinger} (i.e.\ $\hat\vv_{r,r'}=\delta_{r,-r'}\hat\vv/2$). Note that the limit $a\to 0^+$ was only partial and, in particular, it is important to keep $a$ finite in the interaction potential in \Ref{hatv}: taking $a\to 0^+$ amounts to $\hat\vv(p)\to g$ (independent of $p$), which is exactly the limit where the Luttinger model becomes formally equal to the massless Thirring model~\cite{Thirring}. However, as is well-known, this latter limit is delicate and requires a non-trivial multiplicative renormalization~\cite{Wilson}. 

It is important to note that, in the derivation above, one assumes $Q\neq \pi/2$. For $Q=\pi/2$ there are additional terms in the effective Hamiltonian that cannot be ignored and that, presumably, open a gap. This is consistent with $Q=\pi/2$ corresponding to half filling. 

\subsection{Bosonization in 1D}
\label{sec2.3} 
The exact solution of the Luttinger model relies on three mathematical facts: First, the normal ordered fermion densities do not commute (as one would naively expect) but obey the commutator relations
\begin{equation} 
\label{fact1} 
[\hat J_r(p),\hat J_{r'}(p')] = r\delta_{r,r'} p\hat\delta(p+p')
\end{equation} 
where the r.h.s.\ is an example of an {\em anomaly}. Second, the kinetic part of the Luttinger Hamiltonian can be expressed in term of the normal ordered fermion densities, 
\begin{equation}
\label{fact2}
\int_{\Lambda^*}\hat{d}k\, rk :\hat\psi_r^\dag(k)\hat\psi\pdag_r(k)\!: = \pi \int_{\tilde\Lambda^*}\frac{\hat{d}p}{2\pi}\, :\! \hat J_r(p)\hat J_r(-p)\!: , 
\end{equation}  
which is known as {\em Kronig identity}. Third, the fermion field operators can be computed as 
\begin{equation} 
\begin{split} 
\label{fact3} 
\hat\psi_\pm(k) = \lim_{\epsilon\to 0^+}\frac1{2\pi\sqrt{\epsilon}}\int_{-L/2}^{L/2} \ee^{\pm \pi\ii \hat j_\pm(0) x/L}(R_\pm)^{\mp 1} \ee^{\pm \pi\ii \hat j_\pm(0) x/L} \\ \times \exp\left( \pm\int_{\tilde\Lambda^*\setminus\{0\}} \hat dp\frac1{p} \ee^{\ii px-\epsilon|p|}\hat J_\pm(p)  \right)\ee^{-\ii kx}dx
\end{split} 
\end{equation} 
with unitary operators $R_\pm$ obeying $R_+R_-=-R_-R_+$ and 
\begin{equation} 
R_r\hat J_{r'}(p) R_r^{-1} = \hat J_{r'}(p) -r\delta_{r,r'}\delta_{p,0}. 
\end{equation} 
The commutator relations in \Ref{fact1} imply that the linear combinations
\begin{equation}
\label{bosons} 
\begin{split} 
\hat\Pi(p) &= \frac1{\sqrt{2}}\left( -\hat J_+(p) + \hat J_-(p) \right)\\ \hat\Phi(p) &= \frac{1}{\ii p\sqrt{2} }\left(\hat J_+(p) + \hat J_-(p) \right)  \quad \end{split} 
(p\neq 0) 
\end{equation} 
are boson operators with the usual canonical commutator relations (CCR) 
\begin{equation} 
[\hat\Pi(p),\hat\Phi^\dag(p')] = -\ii\hat\delta(p-p'),\quad [\hat\Pi(p),\hat\Pi^\dag(p')] =[\hat\Phi(p),\hat\Phi^\dag(p')] =0
\end{equation} 
with $\hat\Pi^\dag(p)=\hat\Pi(-p)$ and similarly for $\hat\Phi$. Moreover, using \Ref{fact2}, one can express the Luttinger Hamiltonan solely in terms of normalized density operators, and inserting \Ref{bosons} one finds
\begin{equation}
\label{bosonizedLutt} 
H_L = \frac{v_F}{2} \int_{\tilde\Lambda^*\setminus\{0\}}\hat{d}p\, :\! \left( (1-\gamma(p)) \hat \Pi^\dag(p)\hat\Pi(p) +(1+\gamma(p))p^2  \hat \Phi^\dag(p)\hat\Phi(p)  \right)\!: +\ldots
\end{equation}  
where the dots indicate zero mode terms that are $O(1/L)$ and that we suppress, to not clutter our presentation.  The formula in \Ref{bosonizedLutt} shows that the Luttinger Hamiltonian is equivalent to a free boson model that can be diagonalized by a boson Bogoliubov tranformation. Moreover, since \Ref{fact3} expresses the fermion field operators in terms of boson operators, it is possible to compute all fermion correlation functions exactly by analytical methods; see e.g.\ Ref.\ \cite{HSU} for details.

\section{Effective model for a 2D lattice fermion system}
\label{sec3} 
In this section we review a series of papers where we propose~\cite{EL0,EL1} and study~\cite{dWL1,dWL2} an effective model for a 2D lattice fermion system of Hubbard type. In these papers we treat a generalization of the 2D \tV model where we also allow for a next-nearest neighbor hopping term proportional to $t'$ but, for simplicity, we restrict ourselves here to the special case $t'=0$.  

\subsection{Mean field results in 2D}
\label{sec3.1} 
Hartree-Fock theory is a variational method where the groundstate of the interacting system is approximated by the one of non-interacting fermions in an external potential, and this potential is determined such that the variational energy is minimized~\cite{BLS}. In mean field theory one further restricts to variational states that are translationally invariant, which often is adequate for translationally invariant models. For the \tV model, there are three mean field parameters, and one, which we call $\Delta$, is proportional to the magnitude of the density difference between neighboring sites, which is assumed to be constant~\cite{dWL1}. A mean field solution with $\Delta>0$ has the physical interpretation of a charge density wave (CDW) phase: the state is only invariant under translations by two sites, and it is usually insulating.  A normal (N) phase corresponds to a state invariant under all lattice translations, i.e.\ $\Delta=0$, and such a state is typically metallic. It is important to note that mean field theory allows to compare the variational energy of three types of states: First, the pure N state, second, the pure CDW state, and third, a phase separated state where parts of the system are in the N and another part of the system in the CDW state~\cite{LW0,dWL1}. If this third state is obtained it proves that the assumption underlying mean field theory is not correct: the true Hartree-Fock groundstate is not translationally invariant, and this is an indication of unusual physical properties that cannot be accounted for by mean field theory. 
 
\begin{figure}[!b]
\vspace{-0.5cm}
\begin{center}
\includegraphics[width=1.0\textwidth]{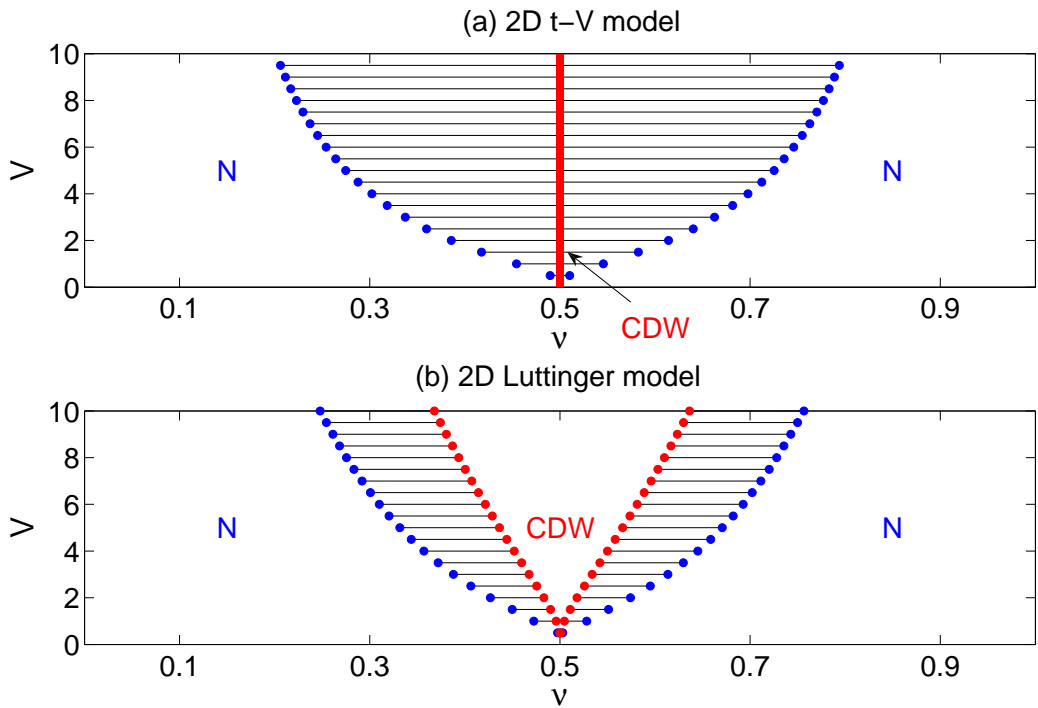}
\end{center}
\vspace{-0.6cm}
\caption{Comparison between the coupling ($V$) vs.\ filling ($\nu$) mean field phase diagrams of the 2D \tV model \textbf{(a)} and the 2D Luttinger model \textbf{(b)} for $t=1$. Shown are the charge-density-wave (CDW) and normal (N) phases as a function of coupling $V$ and filling $\nu$ at zero temperature. The regime  marked by horizontal lines are mixed, i.e.\ neither the CDW nor the N phase is thermodynamically stable. The CDW phase in \textbf{(a)}  exists only at half-filling $\nu=0.5$. The CDW region in \textbf{(b)} corresponds to a partially gapped phase. The parameters used for (b) are $\kappa=0.8$ and $Q/\pi=0.45$.  (Figure reprinted from Ref.\ 4 with kind permission from Springer Science+Business Media.)} 
\label{Fig1}
\end{figure}

Figure~\ref{Fig1}(a) shows the mean field phase diagram for the 2D \tV model as a function of filling $\nu$ and interaction strenght $V/t$, setting $t=1$. As shown, the pure CDW state occurs, but only at strictly half filling, and the N state only in a region quite far away from half filling: in the shaded regions between mean field theory fails. We believe that this failure of mean field theory in a large region away from half filling is due to treating all fermion degrees of freedom in the same way. In the 2D Luttinger model these degrees of freedom are disentangled and thus can be treated by different methods. As shown in Fig.~\ref{Fig1}(b), mean field theory is applicable in a much larger part of the doping regime if used only for the antinodal fermions. As we proposed~\cite{dWL1,dWL2}, the CDW region in Fig.~\ref{Fig1}(b) corresponds to a partially gapped phase where the antinodal fermions are gapped and thus do not contribute to the low energy properties of the system, and this region can be described by the Mattis model.

\subsection{Partial continuum limit of the 2D \tV model} 
We consider the model defined by the Hamiltonian in \Ref{HtV}, but now on a 2D diagonal square lattice with lattice constant $a$ and size $L$, i.e., the lattice sites are $x=(x_1,x_2)$ with $x_{1,2}$ integer multiples of $a$ such that the possible fermion momenta are $\vk=(k_1,k_2)$ with  $k_\pm=(k_1\pm k_2)\sqrt{2}$ half-integer multiples of $2\pi/L$. We  denote the set of all such fermion momenta as $\Lambda^*$. 

The 2D Brillouin zone is a subset of a square (rather than of an interval in 1D),
\begin{equation}
\BZ = \left\{ \vk\in\Lambda^*; \quad -\frac{\pi}{a}<k_{1,2}<\frac{\pi}{a} \right\} ,
\end{equation} 
and, similarly as in 1D, the Hamiltonian in Fourier space is fully characterized by the dispersion relation
\begin{equation}
\label{eps2D} 
\epsilon(\vk) = -2t[\cos(ak_1)+\cos(ak_2)] 
\end{equation} 
and the interaction vertex
\begin{equation} 
v(\vk_1,\ldots,\vk_4) =  \hat u(\vk_1-\vk_2)\sum_{\vn\in\Z^2}\hat\delta^2(\vk_1-\vk_2+\vk_3-\vk_4+(2\pi/a)\vn )
\end{equation} 
with $\hat u(\vp)=a^2V[\cos(ap_1)+\cos(ap_2)]/(2\pi)^2$, i.e., the Hamiltonian can be written as in \Ref{HtV1D} etc.\ but with the obvious replacements $\hat\psi^{(\dag)}(k)\to\hat\psi^{(\dag)}(\vk)$, 
\begin{equation*}
\int \hat{d}k\to \int\hat{d}^2k\equiv \sum_{\vk}\left(\frac{2\pi}{L}\right)^2,\quad \hat\delta(k-k')\to \hat\delta^2(\vk-\vk') = \left(\frac{L}{2\pi}\right)^2\delta_{\vk,\vk'}
\end{equation*} 
etc. 

\begin{figure}[ht!]
\begin{center}
\includegraphics[width=0.8\textwidth]{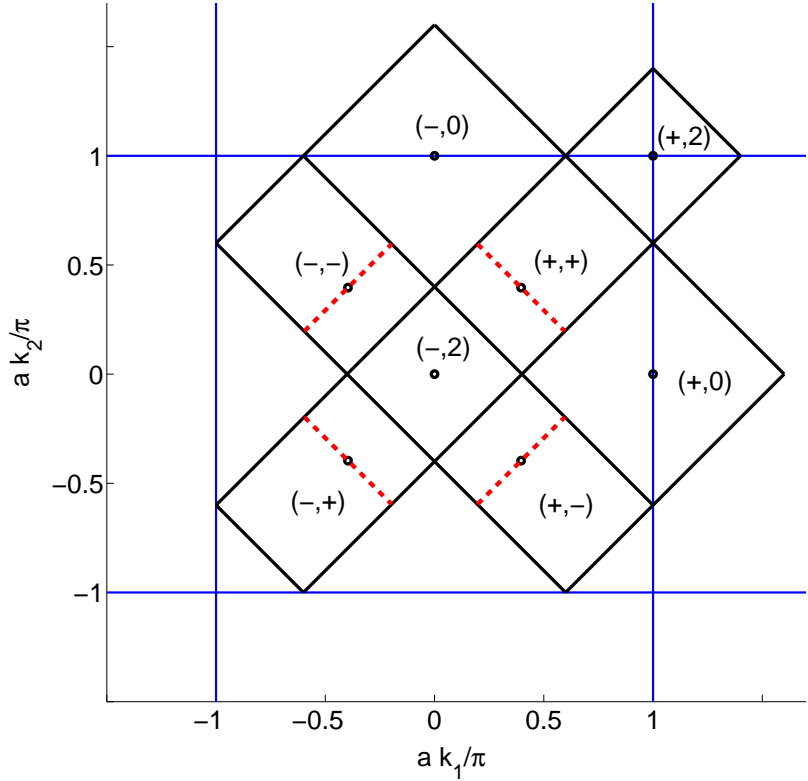}
\end{center}
\caption{Division of the Brillouin zone $\BZ$ in eight regions $\vK_{r,s}+\Lambda^*_{r,s}$ marked as $(r,s)$ for $r=\pm$, $s= 0,\pm ,2$. The eight dots mark the points  $\vK_{r,s}$, and the parameters are $Q=0.45\pi$ and $\kappa=0.8$.  Note that $k_{1,2}$ are only defined up to integer multiples of  $2\pi/a$, and thus the union of these regions cover the $\BZ$  exactly once.  (Figure reprinted from Ref.\ 3 with kind permission from Springer Science+Business Media.)}
\label{FIG3}
\end{figure}

The non-interacting Fermi surface at half filling coincides with the tilted square given by the four line segments  $k_1+sk_2=r\pi/(2a)$, $r,s=\pm$, i.e., $\epsilon(\vk)=0$ on these lines. The midpoints of the sides of this square Fermi surface are $(rQ,rsQ)/a$ with $Q=\pi/2$. We are interested in the model close to half-filling and assume that there is a Fermi surface containing four points 
\begin{equation} 
\label{Krs}
\vK_{r,s}\equiv (rQ,rsQ)/a,\quad r,s=\pm 
\end{equation} 
close to these midpoints, i.e., $Q\approx \pi/2$ is a parameter. Close to these four points, the dispersion relation has a linear behavior,
\begin{equation} 
\label{napprox} 
\epsilon(\vK_{r,s}+\vk) = -4t\cos(Q)+ rv_F k_s +O(|a\vk|^2) ,\quad r,s=\pm 
\end{equation} 
with $v_F = 2\sqrt{2}\sin(Q)ta$. This suggests to approximate the dispersion relation in \Ref{eps2D} by a linear one, keeping only the leading non-trivial term in \Ref{napprox}. This leads to a Fermi surface that includes line segments through the points $\vK_{r,s}$, as indicated by dashed lines in Fig.~\ref{FIG3}. However, there are momentum states close to the non-interacting Fermi surface at half filling where this approximation is qualitatively wrong: the points $\vK_{+,0}\equiv (\pi,0)/a$ and $\vK_{-,0}\equiv (0,\pi)/a$ on this Fermi surface are saddle points of the dispersion relation in \Ref{eps2D}, and 
\begin{equation} 
\label{aapprox}
\epsilon(\vK_{r,0}+\vk) = -rc_F(k_1^2-k_2^2)/2 + O(|a\vk|^3) ,\quad r=\pm 
\end{equation} 
with $c_F=2t a^2$. This hyperbolic behavior of the dispersion relation is important for the tendency of 2D lattice fermion system to develop a gap at half filling~\cite{Schulz}, and it certainly cannot be ignored. There are two more special points, namely $\vK_{+,2}=(\pi,\pi)/a$ and $\vK_{-,2}=(0,0)$, corresponding to a minimum and maximum of the dispersion relation, respectively. However, the regions close to the latter points are far away from the assumed Fermi surface, and we expect that they therefore can be ignored.

This suggests to us to write the 2D \tV model as a model of eight flavors of fermion operators $\hat\psi_{r,s}(\vk)\equiv\hat\psi(\vK_{r,s}+\vk)$, $r=\pm$ and $s=0,\pm,2$, similarly as in 1D. For that we use a partition of the Brillouin zone in eight regions $\vK_{r,s}+\Lambda_{r,s}^*$ such that 
\begin{equation}
\int_{\BZ}\hat{d}^2k\, f(\vk) = \sum_{r=\pm}\sum_{s=0,\pm,2} \int_{\Lambda^*_{r,s}}\hat{d}^2k\, f(\vK_{r,s}+ \vk), 
\end{equation} 
as indicated in Figure~\ref{FIG3}. As already mentioned, we refer to the fermions with $s=\pm$ as {\em nodal} and the ones with $s=0$ as {\em antinodal}.  It is important to note that this partition depends on a parameter $\kappa$ in the range $0<\kappa<1$: the antinodal regions are 
\begin{equation} 
\Lambda_{r,0}^* =\Lambda_0^*  \equiv\left\{\vk\in\Lambda^*;\quad -\kappa\frac{\sqrt2\pi}{a}<k_\pm<\kappa\frac{\sqrt2\pi}{a}\right\}
\end{equation} 
with $0<\kappa<1$, and this also determines the widths of the nodal regions: $\Lambda_{r,s=\pm}$ only contains momenta $\vk$ such that $-\pi/\ta<k_{-s}<\pi/\ta$ with 
\begin{equation}
\label{ta} 
\ta \equiv \frac{\sqrt{2}a}{1-\kappa}. 
\end{equation} 
As discussed in more detail below, $\ta$ is important since it serves as UV cutoff for the  nodal fermions. 

We now modify the 2D \tV model, written as a eight flavor model, similarly as in 1D: First, the nodal- and antinodal dispersion relations are modified by ignoring  the $O(|a\vk|^n)$-terms in \Ref{napprox} ($n=2$) and \Ref{aapprox} ($n=3$), respectively. Second, the interaction vertices are modified by using the obvious generalization of \Ref{1Dvertex} and \Ref{app2} to 2D. Third, the restriction on the nodal momenta orthogonal to the assumed Fermi surface is removed, i.e., 
\begin{equation} 
\label{QFT2D} 
\Lambda^*_{r,s=\pm}\to \Lambda_s^*\equiv\left\{\vk\in\Lambda^*;\quad -\frac{\pi}{\ta}<k_{-s}<\frac{\pi}{\ta} \right\} 
\end{equation} 
with $\ta$ in \Ref{ta}, and, at the same time, all terms in the Hamiltonian involving fermion flavors with $s=2$ are ignored. Note that this third change only modifies fermion degrees far away from the assumed Fermi surface. Similarly as in 1D, it is the change in \Ref{QFT2D} that leads to a QFT model, and this model is mathematically well-defined if this change is done after normal ordering~\cite{EL1}. By lengthy computations one finds a Hamiltonian which, remarkably, only contains interaction terms involving densities
\begin{equation} 
\!: \hat\rho_{r,s}(\vp)\!: \,  \equiv \int_{\Lambda_s^*} \hat{d}^2k_1\int_{\Lambda_s^*} \hat{d}^2k_2\, :\! \hat\psi^\dag_{r,s}(\vk_1)\hat\psi^{\pdag}_{r,s}(\vk_2):\! \hat\delta^2(\vk_1-\vk_2+\vp)
\end{equation} 
for $r=\pm$, $s=0,\pm$. A final change is needed to obtain the 2D Luttinger model: for reasons explained below, we replace in the Hamiltonian
\begin{equation} 
\label{lastchange} 
\begin{split} 
\!: \hat\rho_{r,s=\pm}(\vp)\!: \, \to 
\hat J_{r,s}(\vp) \equiv \int_{\Lambda_s^*} \hat{d}^2k_1\int_{\Lambda_s^*} \hat{d}^2k_2\, :\! \hat\psi^\dag_{r,s}(\vk_1)\hat\psi^{\pdag}_{r,s}(\vk_2):\! \\ \times \sum_{n\in\Z} \hat\delta^2(\vk_1-\vk_2+\vp+2\pi n\ve_{-s}/\ta)
\end{split} 
\end{equation} 
with $\ve_\pm=(1,\pm 1)/\sqrt 2$.  Note that this last change amounts to adding umklapp terms to the nodal densities. 

The changes described above lead to~\cite{EL1} 
\begin{equation} 
H_{tV}\to \cE_0 + H_n+H_a+H_{na}
\end{equation} 
with the nodal part of the Hamiltonian 
\begin{equation} 
\begin{split} 
H_n = \sum_{r,s=\pm} \int_{\Lambda^*_s}\hat{d}^2k\, v_Frk_s :\!\hat\psi^\dag_{r,s}(\vk)\hat\psi^{\pdag}_{r,s}(\vk)\!: + \int_{\tilde\Lambda^*}\frac{\hat{d}^2p}{(2\pi)^2} \sum_{r,s,r',s'=\pm}\\ \times  g\left( \delta_{s,s'}\delta_{r,-r'} +\delta_{s,-s'} \right)\chi_s(\vp)\chi_{s'}(\vp)  \hat J_{r,s}(-\vp)\hat J_{r',s'}(\vp), \\
\end{split} 
\end{equation} 
the antinodal part
\begin{equation} 
\begin{split} 
H_a= \sum_{r=\pm} \int_{\Lambda^*_0}\hat{d}^2k\, (-rc_Fk_+k_--\mu_0)\!:\hat\psi^\dag_{r,0}(\vk)\hat\psi^{\pdag}_{r,0}(\vk)\!: \\ + \int_{\tilde\Lambda^*}\frac{\hat{d}^2p}{(2\pi)^2} \tilde g :\!\hat\rho_{+,0}(-\vp)\!: :\!\hat\rho_{-,0}(\vp)\!: , 
\end{split} 
\end{equation} 
and the nodal-antinodal interactions
\begin{equation}
H_{na} =  \int_{\tilde\Lambda^*}\frac{\hat{d}^2p}{(2\pi)^2}\sum_{r,r',s=\pm} \tilde g\chi_s(\vp):\!\hat\rho_{r,0}(-\vp)\!: \hat J_{r',s}(\vp),
\end{equation} 
with the coupling constants $g=2Va^2\sin^2(Q)$, $\tilde g=2Va^2$, and the cutoff functions $\chi_s(\vp)=\theta(\pi/\ta-|p_{-s}|)\theta(\kappa \pi/(\sqrt{2}a)-|p_s|)$; see Refs.\ \cite{EL1,dWL1} for further details, including explicit formulas for the constants $\cE_0$ and $\mu_0$, and how filling $\nu$ depends on $Q$ and $\kappa$. The Hamiltonian $H=H_n+H_a+H_{na}$ makes mathematically precise a special case of the one in \Ref{2DLutt}. 

Similarly in 1D, one assumes $Q\neq \pi/2$ in the above derivation, and for $Q=\pi/2$ additional nodal interaction terms appear that cannot be bosonized in a simple manner and that, presumably, open a nodal gap. This is consistent with $Q=\pi/2$ corresponding to half filling. 

\subsection{Bosonization in 2D}
The 2D analogues of the identities in \Ref{fact1} and \Ref{fact2} are~\cite{dWL2}
\begin{equation} 
\label{fact12D}
\Bigl[\hat{J}_{r,s}(\vp), \hat{J}_{r',s'}(\vp')\Bigr]= r \delta_{r,r'}\delta_{s,s'} \frac{2\pi p_s }{\ta}\sum_{n\in\mathbb{Z}}\hat\delta^2(\vp+\vp'-2\pi n\ve_{-s}/\ta)
\end{equation}
\begin{equation}
\label{fact22D}
\int_{\Lambda_s^*}\hat{d}^2k\, rk_s :\!\hat\psi^\dag_{r,s}(\vk)\hat\psi\pdag_{r,s}(\vk)\!:\; = \ta\pi \int_{\tilde\Lambda} \frac{\hat{d}^2p}{(2\pi)^2}:\! \hat{J}_{r,s}(-\vp) \hat{J}_{r,s}(\vp)\!: . 
\end{equation}
with $r,s,r's,'=\pm$; see Ref.\ \cite{EL1}, Proposition~2.1 for a mathematically precise formulation. Moreover, there exists a generalization of \Ref{fact3} to 2D; see Ref.\ \cite{EL1}, Proposition~2.7. 

Similarly as in 1D, the first of these identities implies that the operators
\begin{equation}
\begin{split} 
\hat\Pi_s(\vp) &= \sqrt{\frac{\ta}{4\pi}}\left(-\hat{J}_{+,s}(\vp) + \hat{J}_{-,s}(\vp) \right)\\  \hat\Phi_s(\vp) &= \sqrt{\frac{\ta}{4\pi}}\frac1{\ii p_s}\left(\hat{J}_{+,s}(\vp) + \hat{J}_{-,s}(\vp) \right)
\end{split} \quad (p_s\neq 0)
\end{equation} 
are boson operators with the usual CCR, and using these operators, the nodal- and mixed parts of the 2D Luttinger Hamiltonian can be written as~\cite{EL1}
\begin{equation} 
\begin{split} 
H_n = \frac{v_F}{2} \sum_{s=\pm}\int_{\tilde\Lambda^*}\hat{d}^2p:\! \Biggl( \left( 1-\gamma_s(\vp) \right)\Pi^\dag_s(\vp) \Pi^{\pdag}_s(\vp) + \left( 1+\gamma_s(\vp) \right)p_s^2 \Phi^\dag_s(\vp) \Phi^{\pdag}_s(\vp)\\ + \gamma(\vp)p_+p_- 
\Phi^\dag_s(\vp) \Phi^{\pdag}_s(\vp) \Biggr)\!:  +\ldots
\end{split} 
\end{equation} 
and
\begin{equation} 
H_{na} =  \int_{\tilde\Lambda^*}\frac{\hat{d}^2p}{(2\pi)^2}\sum_{r,s=\pm} \sqrt{\frac{4\pi}{\ta}}\ii p_s \tilde g\chi_s(\vp):\!\hat\rho_{r,0}(-\vp)\!: \hat\Phi_s(\vp),
\end{equation} 
with $\gamma_s(\vp)=\gamma\chi_s(\vp)$, $\gamma(\vp)=\gamma\chi_+(\vp)\chi_-(\vp)$, and $\gamma=V(1-\kappa)\sin(Q)/(2\pi t)$; as before, the dots indicate zero mode terms that are $O(1/L)$. 

\subsection{Relation to the Mattis model} 
As described in the previous section, the 2D Luttinger Hamiltonian can be mapped exactly to a Hamiltonian describing non-interacting bosons coupled linearly to interacting antinodal fermions. It is possible to integrate out these nodal bosons exactly and thus obtain an effective model for the antinodal fermions only. In Ref.~\cite{dWL1} we studied this latter model by conventional approximation methods including mean field theory, and we found a non-trivial region away from half filling where the antinodal fermions are gapped. We assume that the gapped fermions do not contribute to the low energy properties of the system, and we therefore proposed that the Mattis model describes the low energy physics of the 2D Luttinger model in this partially gapped phase. 

\subsection{Solution of the Mattis model} 
Since the nodal part of the 2D Luttinger Hamiltonian is quadratic in boson operators, it can be diagonalized by a boson Bogoliubov transformation. From this, we can compute the ground state- and free energy of the nodal (Mattis) model~\cite{dWL2}. Furthermore, there is a natural generalization of the boson-fermion correspondence formula in \Ref{fact3} to the two-dimensional case~\cite{dWL2}. This formula allows to compute all fermion correlation functions exactly using standard results for free bosons. For example, we found that the fermion two-point function has algebraic decay for intermediate length scales, and with non-trivial exponents that depend on the coupling constants.

The result for the two-point function may be interpreted as a signature of Luttinger-liquid behavior in the Mattis model. We emphasize however that this does not automatically imply Luttinger-liquid behavior in the 2D \tV model of lattice fermions, even if the antinodal fermions are gapped. For this to hold true, one first needs to investigate the approximations introduced when deriving the 2D Luttinger model from the lattice system.

\section{Extensions of the Mattis model}
\label{sec4} 
There are several ways to obtain new exactly solvable models of interacting fermions by extending the Mattis model. 
Below we shortly describe two such extensions~\cite{dWL3,dWL4}.
\subsection{A 2+1D quantum gauge theory with interacting fermions}
\label{sec4.1} 
In Ref.~\cite{dWL3}, we consider the quantum gauge theory model obtained by minimally coupling the fermions in the Mattis model to a two-dimensional dynamical electromagnetic field. The gauged model shares many features with the so-called Schwinger model~\cite{Schwinger} of (1+1)D quantum electrodynamics with massless Dirac fermions; see also Refs.~\cite{Manton,GrosseLangmannRaschhofer}. The anomaly in \Ref{fact12D} and the requirement of gauge invariance on the quantum level leads to a bare mass term proportional to the electric charge for the electromagnetic field, similarly as for the Schwinger model. 

The Hamiltonian of the gauged model can be bosonized and subsequently diagonalized; one finds two gapped, or massive, boson modes and one gapless mode. We also computed the linear response of the magnetic field to an external current and found that there is a Meissner effect. 

\subsection{Partial continuum limit of the 2D Hubbard model}
\label{sec4.2}
Our work on the 2D \tV model of spinless fermions can be extended to the Hubbard model (see \Ref{HtU}). In Ref.~\cite{dWL4}, we applied the partial continuum limit to the 2D Hubbard model on a square lattice. This again leads to an effective QFT model of nodal fermions coupled to antinodal fermions. The nodal fermions can be bosonized using either abelian- or non-abelian~\cite{Witten} methods, and in the latter case we obtain a natural 2D analogue of a Wess-Zumino-Witten model. Furthermore, by a specific truncation of the nodal part of the effective Hamiltonian, we obtain a spinfull variant of the Mattis model that is exactly solvable. The fundamental excitations of this model separate into independent spin- and charge degrees of freedom. 

\section{Discussion}
\label{sec5} 
The \HTC problem has been an outstanding challenge in theoretical physics for many years; see for example Ref.~\cite{feature} for recent commentaries by leading experts on the status of this field. One important hypothesis underlying much of the work on the \HTC problem is that {\em the 2D Hubbard model is an adequate prototype model for the cuprates}. From a theorist's point of view this hypothesis is highly appealing due to the (deceptively) simple and aesthetic form of the model. Furthermore, the model is considered "adequate" in the sense that, \nr{i} it is possible to derive it using physical arguments from a more realistic description and,  \nr{ii} it contains enough physics to capture the qualitative properties of the cuprates, while better quantitative agreement can be obtained by straightforward extensions of the model. Still, while there seem to be little doubts about \nr{i} and \nr{ii}, these criteria are not enough; an adequate model also requires that, \nr{iii} it should be amenable to reliable computations and thus allow for experimental predictions. This criterion, which is obviously the most important one, remains to be fulfilled.   

The results reviewed in this paper are part of a program aiming at finding a better balance between \nr{i}--\nr{iii} by modifying a 2D Hubbard-like lattice fermion model. Our emphasis has been on modifications that lead to exactly solvable models since, for such models, \nr{iii} is always true. However, it is important to note that, even though we have proposed a derivation of such a model from a Hubbard-like model, we cannot claim to have shown that \nr{i} is fulfilled: our derivation contains steps that need to be substantiated by mathematical arguments (these steps are spelled out in Refs.~\cite{EL1} and \cite{dWL2}). We also stress that, up to now, \nr{ii} remains open: we have not systematically investigated whether the Mattis model, or its various extensions, can describe the cuprates. Much remains to be done in our program.  

\section*{Acknowledgements}
This work was supported by the G\"oran Gustafsson Foundation and the Swedish Research Council (VR)  under contract no. 621-2010-3708. We thank Farrokh Atai for carefully reading the manuscript.

\end{document}